\documentclass[aps,prd,amsmath,tightenlines,amsfonts,preprintnumbers,nofootinbib,amssymb]{revtex4}
\usepackage{graphicx}

\newcommand{\figref}[1]{Fig.~\ref{fig:#1}}

\newcommand{\tableref}[1]{Table~\ref{table:#1}}

\begin{document}

\preprint{FERMILAB-CONF-13-304-T}

\title{Di-jet resonances at future hadron colliders: A Snowmass whitepaper}
\author{Felix Yu}
\email{felixyu@fnal.gov}
\affiliation{Theoretical Physics Department, Fermilab, 
Batavia, IL 60510, USA}

\begin{abstract}
I investigate the sensitivity of future hadron colliders to di-jet
resonances arising from $Z^\prime$ or coloron models.  The projected
discovery potential and exclusion limits for these resonances is
presented in the coupling vs. mass plane, which highlights both the
increased mass reach from higher energy machines as well as the
improved coupling sensivity from larger luminosity.
\end{abstract}

\maketitle

If history holds, searches for dijet resonances will be the first
beyond the standard model (BSM) analyses performed at future hadron
colliders~\cite{Aad:2010ae, Khachatryan:2010jd}.  We are thus
motivated to understand the discovery reach and exclusion sensitivity
for these resonances at such machines.  Besides being useful probes
for new physics, we note such searches are also useful standard
candles for understanding and calibrating new detectors in an
unfamiliar and exciting collider environment.

Of the possible spin and initial state combinations to produce a dijet
resonance at hadron colliders, the most theoretically straightforward
constructions are new color singlet or color octet vector resonances
arising from $q \bar{q}$ annihilation~\cite{Dobrescu:2013cmh}.  This
is driven by requiring couplings that are renormalizable and
flavor-universal, which reflect a desire for models with unsuppressed
production rates that simultaneously satisfy flavor bounds.

There are two parameters that characterize a purely leptophobic,
flavor-universal dijet resonance: mass and coupling.  Following
Ref.~\cite{Dobrescu:2013cmh}, the coupling--mass plane provides a
unified presentation of experimental limits, allowing ready
interpretation of different experimental searches performed with
different integrated luminosities and at different $\sqrt{s}$.  In
particular, when considering future hadron colliders, the
coupling--mass plane highlights the possible reach in both mass from
higher $\sqrt{s}$ and coupling from larger luminosity.  

As in Ref.~\cite{Dobrescu:2013cmh}, we consider color singlet
($Z^\prime_B$) and color octet ($G'$, coloron) vector resonances.  The
$Z^\prime_B$ model is equivalent to a gauged $U(1)$ baryon number,
where additional colored fermions are added to cancel
anomalies.\footnote{Details of the $Z^\prime_B$ model construction can
  be found in Ref.~\cite{Dobrescu:2013cmh} and its references.}  The
coupling $g_B$ is normalized to the baryon number charge $1/3$ of the
standard model (SM) quarks, giving
\begin{equation}
\mathcal{L} \supset \frac{g_B}{6} Z^\prime_{B \mu} \bar{q} \gamma^\mu q \ .
\end{equation}

\begin{figure*}[t]
\begin{center}
\includegraphics[width=0.47\textwidth, angle=0]{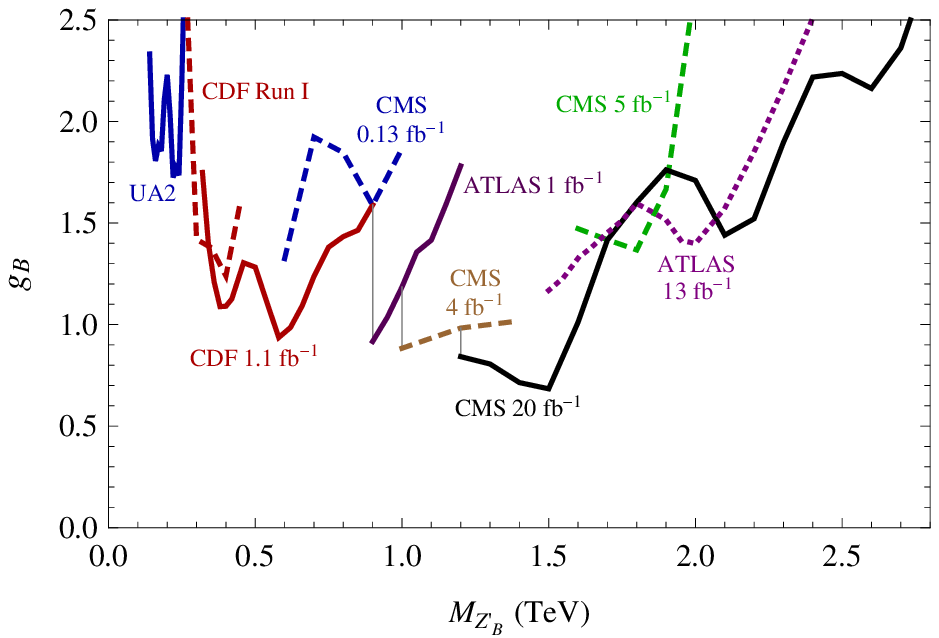}
\hspace{0.2cm}
\includegraphics[width=0.47\textwidth, angle=0]{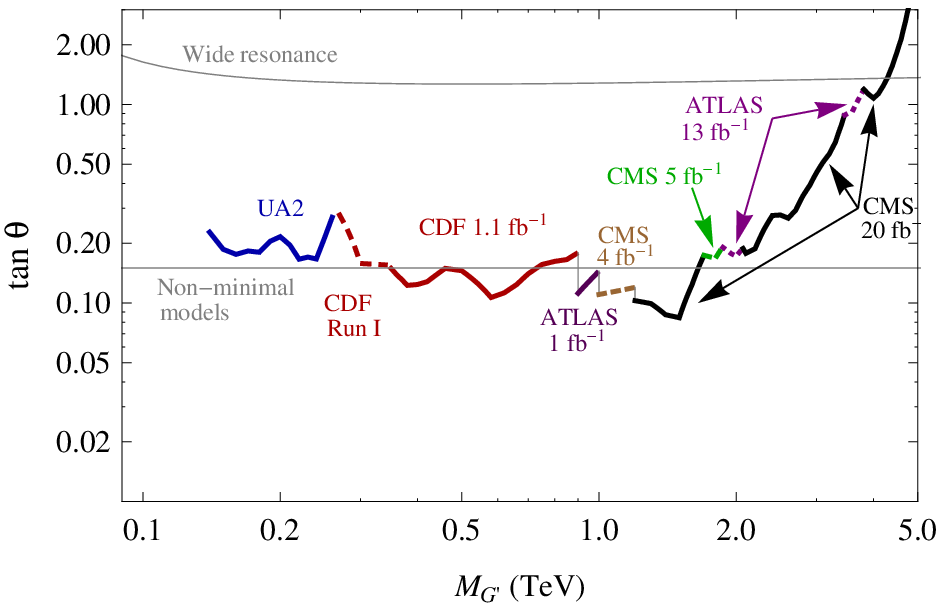}
\caption{Leading experimental limits in the ({\bf left}) coupling $g_B$
  versus mass $M_{Z^\prime_B}$ plane for $Z^\prime_B$ resonances and
  ({\bf right}) coupling $\tan \theta$ versus mass $M_{G^\prime}$ plane for
  coloron resonances.  Values above each line are excluded at the 95\%
  C.L.  Plots from Ref.~\cite{Dobrescu:2013cmh}.}
\label{fig:Limitsnow}
\end{center}
\end{figure*}

A massive color octet vector resonance arises in coloron models, where
an extended gauge symmetry group of $SU(3) \times SU(3)$ breaks to the
diagonal $SU(3)_c$ subgroup, which we identify as the SM color gauge
group~\cite{Bai:2010dj}.  Assuming all SM quarks are universally
charged under one of the parent $SU(3)$ gauge groups, the interaction
term of the coloron to the SM quarks is
\begin{equation}
\mathcal{L} \supset g_s \tan \theta \bar{q} \gamma^\mu T^a G^{\prime
  a}_\mu q \ ,
\end{equation}
where $g_s$ is the strong coupling constant.  The current status of
dijet resonance searches in the coupling--mass plane is displayed
in~\figref{Limitsnow} for ($M_{Z^\prime_b}$, $g_B$) and
($M_{G^\prime}$, $\tan \theta$): the leading dijet search limits are
from Refs.~\cite{Alitti:1993pn, Abe:1997hm, Aaltonen:2008dn,
  CMS:2012cza, Aad:2011fq, Chatrchyan:2013qha, CMS:2012yf,
  ATLAS:2012qjz, CMS:kxa} and the plots are reproduced from
Ref.~\cite{Dobrescu:2013cmh}.

We investigate the future sensitivity to dijet resonances at the 14
TeV LHC, a possible 33 TeV $pp$ collider, and a future 100 TeV $pp$
collider.  We use MadGraph 5 v.1.5.7~\cite{Alwall:2011uj} with the
CTEQ6L1 PDFs~\cite{Pumplin:2002vw} to generate signal events for both
the $Z^\prime_B$ and the $G'$ resonances.  These events are passed
through \textsc{Pythia} v6.4.20~\cite{Sjostrand:2006za} for showering
and hadronization, and then through PGS v4~\cite{PGS4} for basic
detector simulation.

\begin{figure*}[t]
\begin{center}
\includegraphics[width=0.30\textwidth, angle=0]{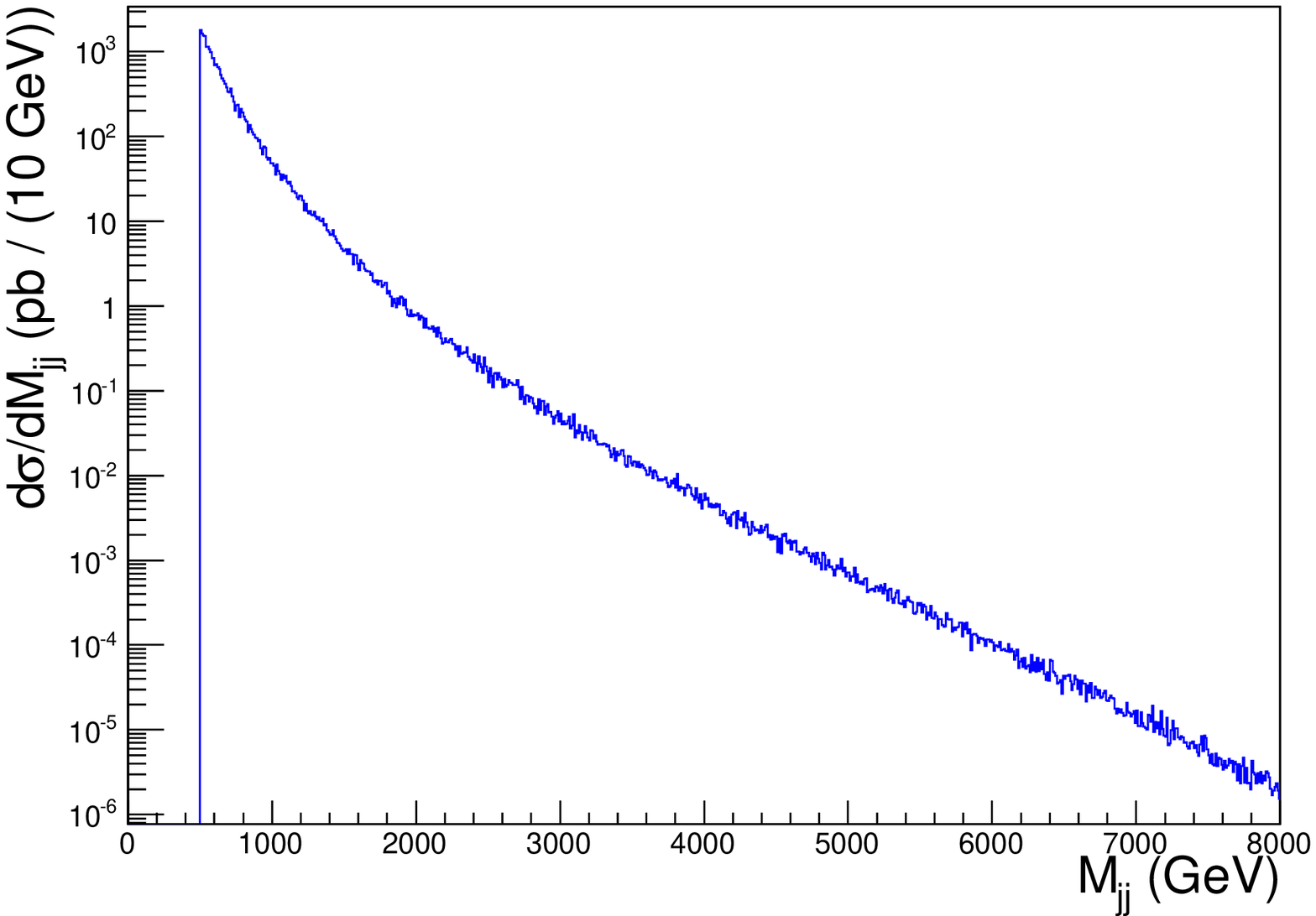}
\hspace{0.2cm}
\includegraphics[width=0.30\textwidth, angle=0]{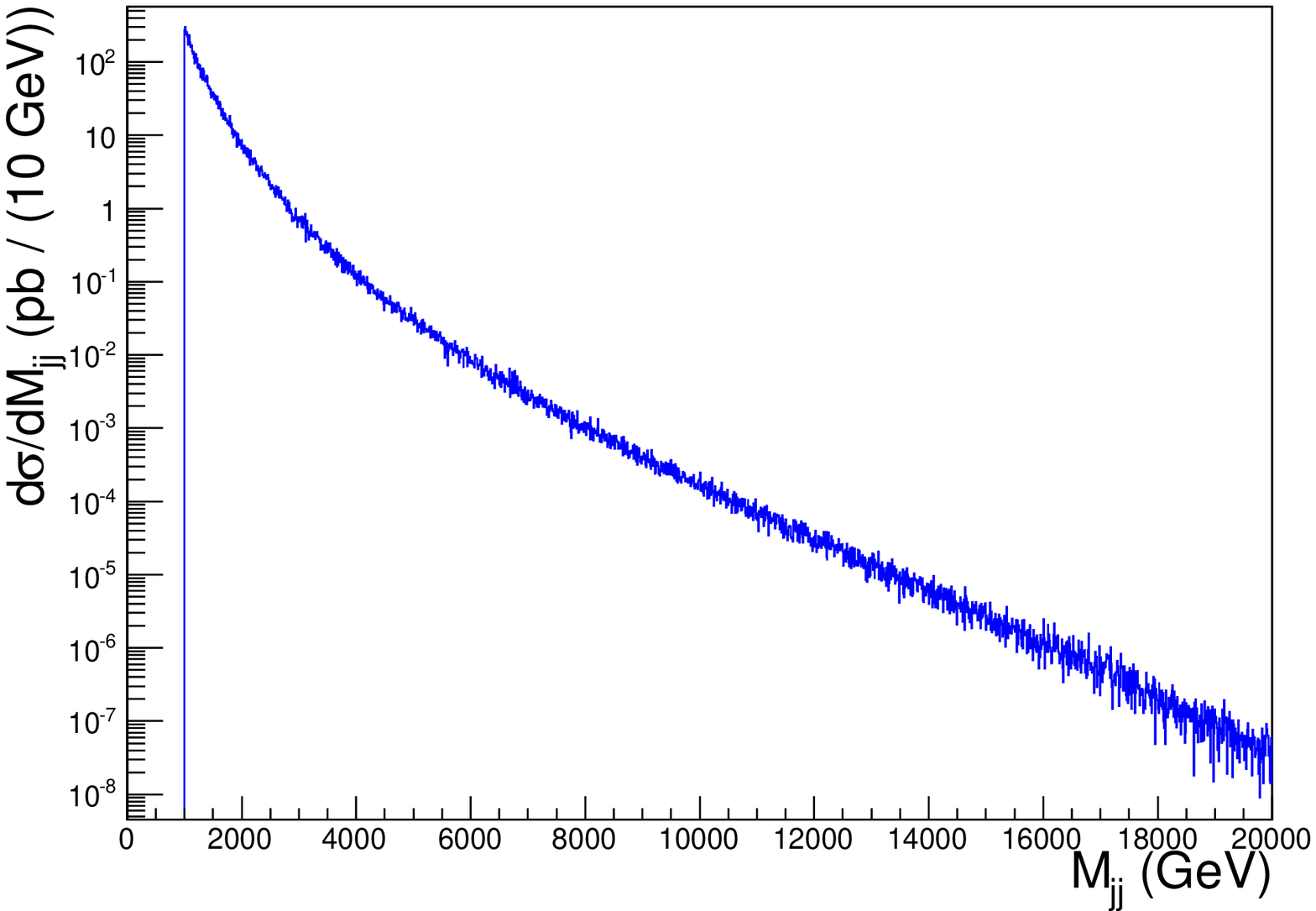}
\hspace{0.2cm}
\includegraphics[width=0.30\textwidth, angle=0]{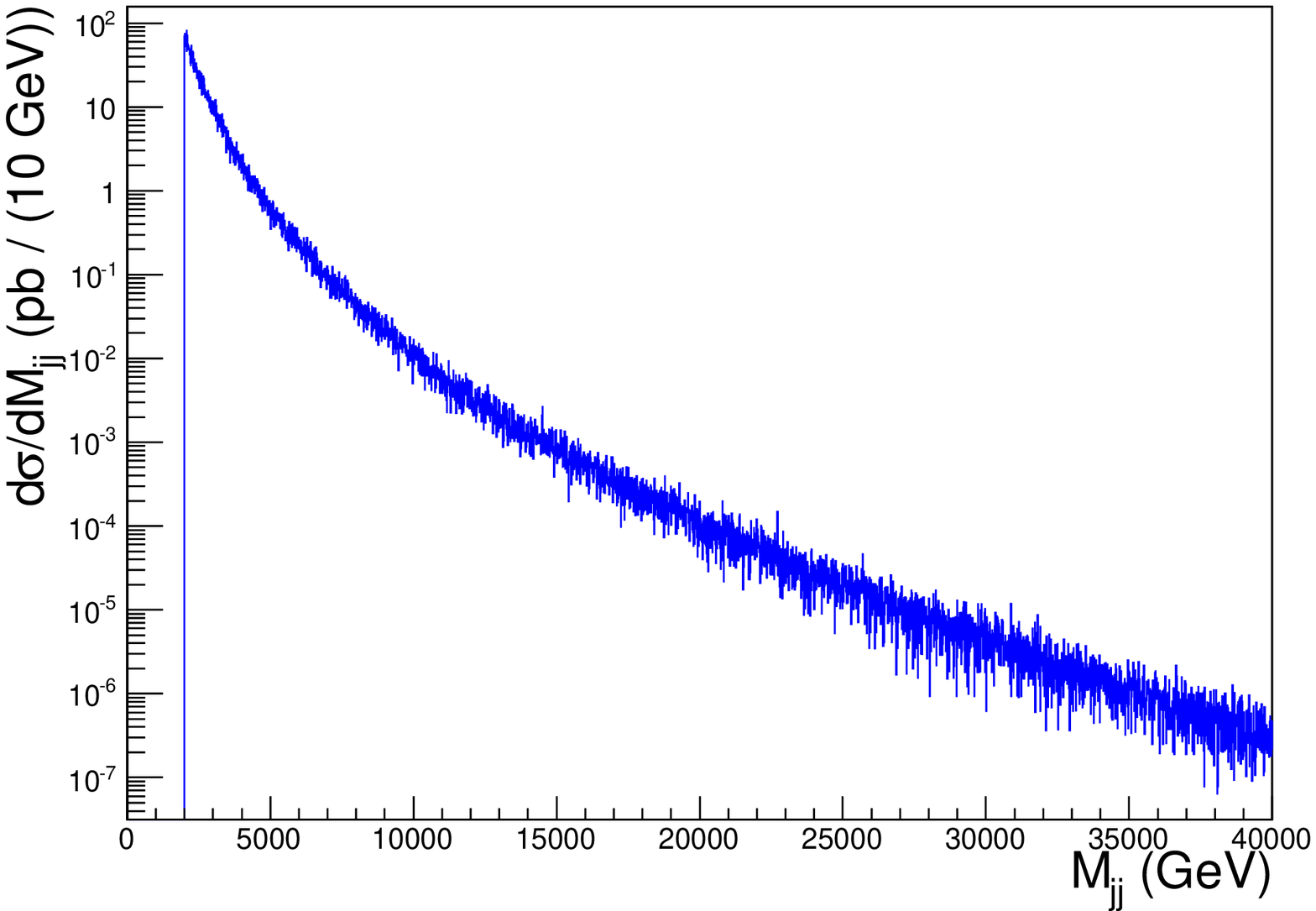}
\caption{Background samples generated for QCD at ({\bf left}) 14 TeV,
  ({\bf middle}) 33 TeV, ({\bf right}) 100 TeV as described in the
  text.}
\label{fig:QCD}
\end{center}
\end{figure*}

The QCD background is also generated in MadGraph 5 with CTEQ6L1 PDFs
and interfaced with \textsc{Pythia} as an MLM~\cite{Mangano:2002ea}
matched sample of two-jet and three-jet events in separate bins of
leading $p_T$, following the prescription of Refs.~\cite{Esen:2006oxa,
  Gumus:2006mxa}.  The events are then clustered by \textsc{FastJet}
v.3.0.2~\cite{Cacciari:2011ma} using the anti-$k_T$
algorithm~\cite{Cacciari:2008gp} with distance parameter $R = 0.5$.
The $p_T$ bins for background generation are listed
in~\tableref{pTbins}.  We then adopt the same analysis cuts
as~\cite{CMS:kxa} to form the dijet mass spectrum.  We show our
results for the background sample in~\figref{QCD}.

\begin{table}[t]
\renewcommand{\arraystretch}{1.2}
\begin{tabular}{|c|c|c|c||c|c|c|c|}
\hline
$p_T$ bin & 14 TeV & 33 TeV & 100 TeV & $p_T$ bin & 14 TeV & 33 TeV & 100 TeV \\
\hline
 1 & $0.100-0.150$ & $0.200-0.300$ & $0.500-0.650$ & 13 & $1.60-1.80$   & $2.75-3.10$ & $4.00-4.75$ \\
 2 & $0.150-0.200$ & $0.300-0.400$ & $0.650-0.800$ & 14 & $1.80-2.00$   & $3.10-3.50$ & $4.75-5.50$ \\
 3 & $0.200-0.250$ & $0.400-0.550$ & $0.800-1.00$ & 15 & $2.00-2.25$   & $3.50-4.00$ & $5.50-6.25$ \\
 4 & $0.250-0.325$ & $0.550-0.700$ & $1.00-1.30$ & 16 & $2.25-2.50$   & $4.00-4.50$ & $6.25-7.00$ \\
 5 & $0.325-0.400$ & $0.700-0.850$ & $1.30-1.55$ & 17 & $2.50-2.80$   & $4.50-5.00$ & $7.00-8.50$ \\
 6 & $0.400-0.500$ & $0.850-1.00$ & $1.55-1.80$ & 18 & $2.80-3.00$   & $5.00-6.00$ & $8.50-10.0$ \\
 7 & $0.500-0.650$ & $1.00-1.25$ & $1.80-2.10$ & 19 & $3.00-3.30$   & $6.00-7.00$ & $10.0-12.5$ \\
 8 & $0.650-0.800$ & $1.25-1.50$ & $2.10-2.40$ & 20 & $3.30-3.75$   & $7.00-8.50$ & $12.5-15.0$ \\
 9 & $0.800-1.00$  & $1.50-1.75$ & $2.40-2.70$ & 21 & $3.75-4.10$   & $8.50-10.0$ & $15.0-17.5$ \\
10 & $1.00-1.20$   & $1.75-2.00$ & $2.70-3.00$ & 22 & $4.10-4.50$   & $10.0-11.5$ & $17.5-20.0$ \\
11 & $1.20-1.40$   & $2.00-2.30$ & $3.00-3.50$ & 23 & $4.50-6.00$   & $11.5-13.0$ & $20.0-25.0$ \\
12 & $1.40-1.60$   & $2.30-2.75$ & $3.50-4.00$ & 24 & $6.00+$       & $13.0+$     & $25.0+$ \\
\hline
\end{tabular}
\caption{\label{table:pTbins} Kinematic bins of leading jet $p_T$ used
  in the QCD background production.}
\end{table}

\begin{figure*}[t]
\begin{center}
\includegraphics[width=0.47\textwidth, angle=0]{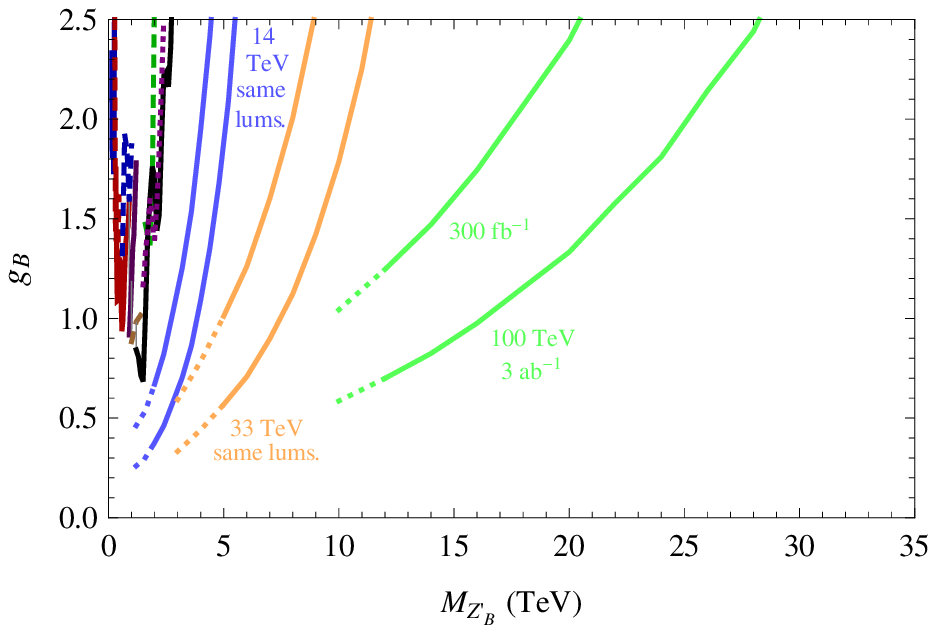}
\hspace{0.2cm}
\includegraphics[width=0.47\textwidth, angle=0]{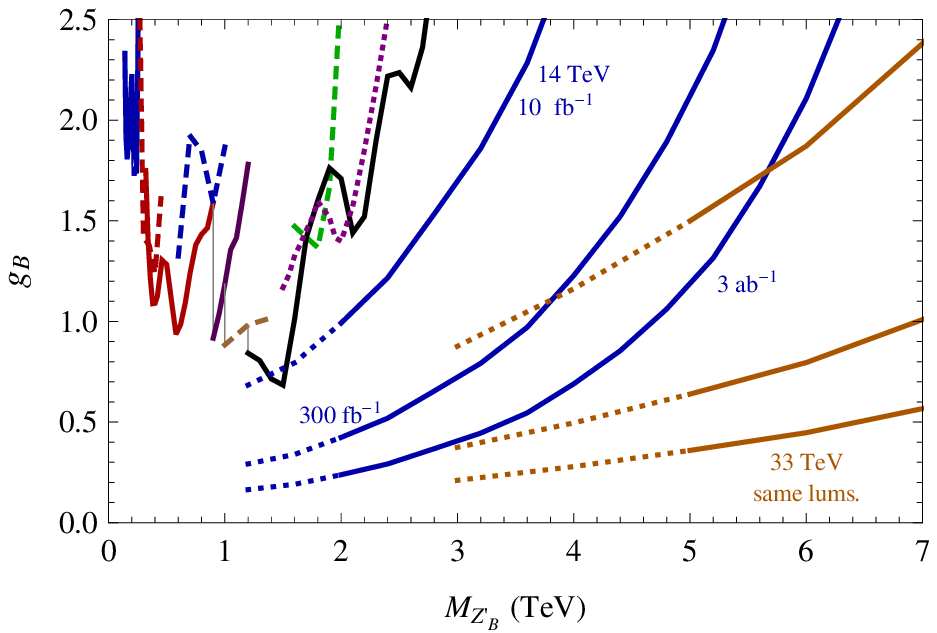} \\
\vspace{0.3cm}
\includegraphics[width=0.47\textwidth, angle=0]{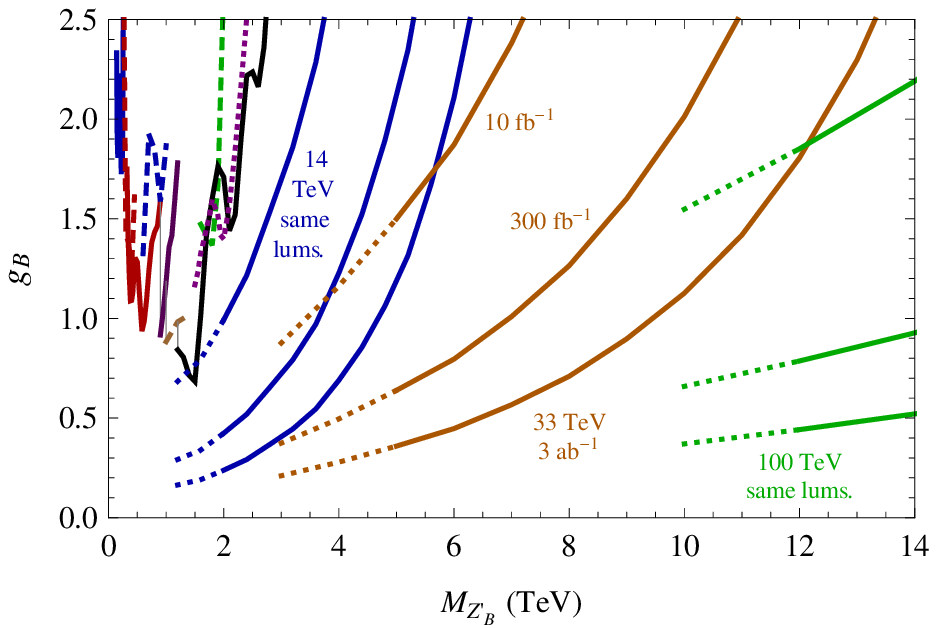}
\hspace{0.2cm}
\includegraphics[width=0.47\textwidth, angle=0]{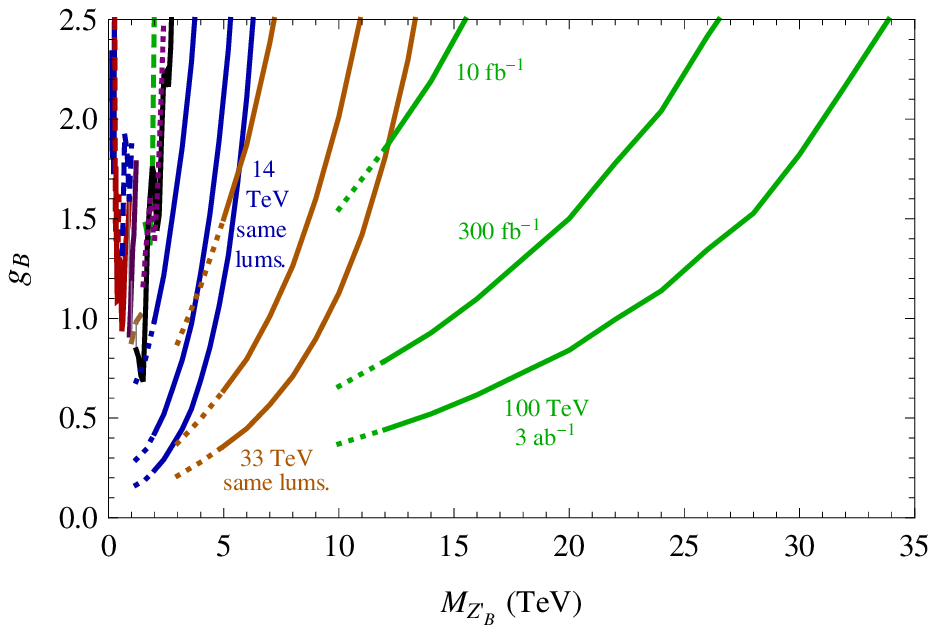}
\caption{{\bf Top left}: Leading experimental limits and projected
  $5\sigma$ discovery sensitivity contours for 14 TeV (light blue
  solid), 33 TeV (light orange solid), and 100 TeV (light green
  solid).  {\bf Top right, bottom left, bottom right}: Leading
  experimental limits and projected 95\% C.L. exclusion contours for
  14 TeV (dark blue solid), 33 TeV (dark brown solid), and 100 TeV
  (dark green solid) $pp$ colliders in the coupling $g_B$ versus mass
  $M_{Z^\prime_B}$ plane for $Z^\prime_B$ resonances.  The dotted
  continuation of each projection line to low masses indicates an
  extrapolation to low multijet trigger thresholds.}
\label{fig:Zplimits}
\end{center}
\end{figure*}

\begin{figure*}[t]
\begin{center}
\includegraphics[width=0.47\textwidth, angle=0]{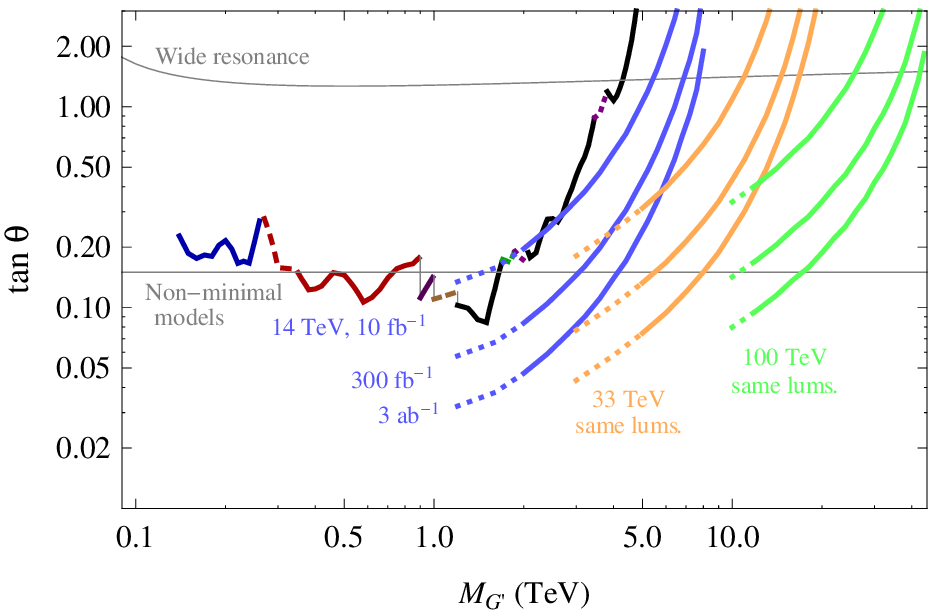}
\hspace{0.2cm}
\includegraphics[width=0.47\textwidth, angle=0]{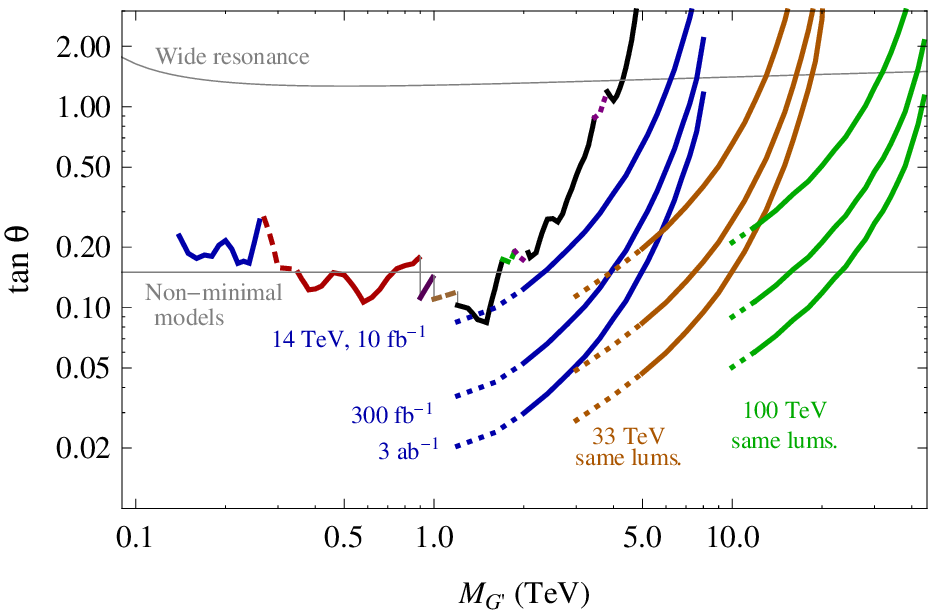}
\caption{Leading experimental limits and ({\bf left}) projected $5
  \sigma$ discovery sensitivity contours and ({\bf right}) $95\%$
  C.L. exclusion contours for 14 TeV (blue solid), 33 TeV (brown
  solid), and 100 TeV (green solid) $pp$ colliders in the coupling
  $\tan \theta$ versus mass $M_{G^\prime}$ plane for coloron
  resonances.  Values of $\tan \theta$ above each line are excluded at
  the 95\% C.L.  The dotted continuation of each projection line to
  low masses indicates an extrapolation to low multijet trigger
  thresholds.}
\label{fig:Colimits}
\end{center}
\end{figure*}

From these background and signal samples, we perform a bump hunt as a
function of the coupling and the resonance mass.  For a given
resonance mass, we use a Crystall Ball fit (see the Appendix of
Ref.~\cite{Dobrescu:2013cmh} to isolate the Gaussian peak feature of
the signal.  These Gaussian parameters dictate the appropriate mass
window of QCD background to compare to signal.  We calculate
statistical significance according to $\sigma = N_S / \sqrt{N_S +
  N_B}$, ignoring possible systematic uncertainties.  We work in the
narrow width approximation, and thus cross sections scale as coupling
squared.  We correspondingly solve for the $5 \sigma$ discovery reach
and the 95\% C.L. exclusion limit.  We show our results
in~\figref{Zplimits} for the $Z^\prime_B$, and in~\figref{Colimits}
for the coloron.  The lowest mass sensitivity in each projection is
dotted to indicate uncertainty about the multijet trigger threshold.
While a complete study of the trigger paths of future hadron colliders
is beyond the scope of this work, it is clear that efforts to drive
multijet trigger thresholds as low as practical is critical to ensure
large sensitivity gaps do not develop as we move to next generation
colliders.

These results show that future hadron colliders are very promising for
extending the reach in both coupling and mass for new dijet
resonances.  The current projection for the 14 TeV LHC with 300
fb$^{-1}$ luminosity indicates $Z^\prime_B$ masses as heavy as 4.5
(5.3) TeV or as weakly coupled as $g_B \sim 0.65$ ($0.4$) could be
discovered (excluded).  A high luminosity run of the LHC with 3
ab$^{-1}$ would extend the discovery reach (exclusion limits) to
masses as heavy as $5.5$ ($6.1$) TeV and couplings as small as $g_B
\sim 0.35$ ($0.2$).  For a 33 TeV or a 100 TeV collider with 3
ab$^{-1}$ integrated luminosity, the $Z^\prime_B$ mass reach for
discovery (exclusion) extends to $11.5$ ($13$) TeV and $28$ ($34$) TeV,
respectively.

The discovery and exclusion prospects for colorons are similarly
impressive.  The 14 TeV LHC with 300 fb$^{-1}$ has discovery potential
(exclusion sensitivity) for colorons as heavy as 6.5 (7.5) TeV or
couplings as small as $\tan \theta \sim 0.08$ ($0.04$).  With 3
ab$^{-1}$ of integrated luminosity, this improves to 7.5 (8.5) TeV in
mass reach and 0.045 (0.030) in $\tan \beta$ for discovery
(exclusion).  With a $\sqrt{s} = 33$ TeV or $\sqrt{s} = 100$ TeV
collider and 3 ab$^{-1}$ dataset, colorons as heavy as 16 (18) TeV and
40 (44) TeV can be discovered (excluded), respectively.  These
fantastic discovery prospects lend strong support for the continued
LHC effort in searching for new physics.  We are also excited by the
improved sensitivity after an LHC luminosity upgrade as well as the
truly impressive discovery reach by future hadron colliders with
higher center of mass energies.

\acknowledgments 
FY would like to acknowledge useful discussions with Ciaran Williams.
Fermilab is operated by the Fermi Research Alliance, LLC under
Contract No.~De-AC02-07CH11359 with the United States Department of
Energy.

\end{document}